\documentclass[twocolumn]{aastex63}

\usepackage{paralist}
\usepackage{pifont}
\usepackage{dblfloatfix}
\usepackage{bm}
\usepackage{hyperref}
\usepackage{amsmath}

\usepackage{graphicx}
\usepackage{dcolumn}
\usepackage{natbib}
\usepackage{color}
\usepackage{amssymb}
\usepackage{tabularx}
\usepackage{yfonts}
\usepackage[rightcaption]{sidecap}

\newcommand{\chg}[1]{{\color{black} #1}}

\newcommand{\jcr}{\tilde{J}_{\rm cr}}

\usepackage{lineno}

\shorttitle{High-CR-Current Streaming Instability}
\shortauthors{Lichko, Caprioli, Schroer \& Gupta}

\begin{document}

\title{Understanding Streaming Instabilities in the Limit of High Cosmic Ray Current Density}

\correspondingauthor{Emily Lichko}
\email{emily.r.lichko.civ@us.navy.mil}

\author[0000-0003-1945-8460]{Emily Lichko}
\affiliation{Department of Astronomy and Astrophysics, The University of Chicago, IL 60637, USA}
\affiliation{Plasma Physics Division, Naval Research Laboratory, Washington, DC 20375, USA}
\author[0000-0003-0939-8775]{Damiano Caprioli}
\affiliation{Department of Astronomy and Astrophysics, The University of Chicago, IL 60637, USA}
\affiliation{Enrico Fermi Institute, The University of Chicago, Chicago, IL 60637, USA}

\author[0000-0002-4273-9896]{Benedikt Schroer}
\affiliation{Department of Astronomy and Astrophysics, The University of Chicago, IL 60637, USA}
\author[0000-0002-1030-8012]{Siddhartha Gupta}
\affiliation{Department of Astronomy and Astrophysics, The University of Chicago, IL 60637, USA}
\affiliation{Department of Astrophysical Sciences, Princeton University, 4 Ivy Ln., Princeton, NJ 08544, USA}

\begin{abstract}
A critical component of particle acceleration in astrophysical shocks is the non-resonant (Bell) instability, where the streaming of cosmic rays (CRs) leads to the amplification of magnetic fields necessary to scatter particles. 
In this work we use kinetic particle-in-cells simulations to investigate the high-CR current regime, where the typical assumptions underlying the Bell instability break down. 
Despite being more strongly driven, significantly less magnetic field amplification is observed compared to low-current cases, an effect due to the anisotropic heating that occurs in this regime. 
We also find that electron-scale modes, despite being fastest growing, mostly lead to moderate electron heating and do not affect the late evolution or saturation of the instability.

\end{abstract}



\section{Introduction\label{sec:introduction}}

Streaming instabilities, where the free energy in the relative drift between two populations of particles transfer energy to the magnetic fields, play a critical role in the dynamics of a wide range of space and astrophysical systems. 
This includes the acceleration of charged particles in heliospheric and astrophysical shocks and the transport of cosmic rays (CRs) in galactic and intergalactic plasmas \citep[e.g.,][]{blasi13, caprioli15p, zweibel17}. 

Arguably the most important instabilities for the energization of CRs are the resonant streaming instability \citep{kulsrud+68, bell78a,zweibel79} and the non-resonant, or Bell, instability \citep{bell04,amato+09}, which are both driven by the super-Alfv\'enic drift of CRs with respect to the background thermal plasma \citep[also see][]{achterberg83,winske+84, weidl+19a}. 
In this work, we will focus on the regime of very high-CR currents, where the assumptions that go into the derivation of resonant and Bell instabilities break down.

The Bell instability essentially hinges on the return current generated in the background plasma by highly energetic particles.
Its linear physics can be derived using a magnetohydrodynamic approximation or a kinetic treatment \citep{bell04, amato+09}. 
In both cases, the derivations usually assume that both the real and imaginary parts of the mode frequencies are much smaller than the non-relativistic ion cyclotron frequency, $|\omega|\ll \omega_{ci}=eB_0/m_ic$, where $B_0$ is the background magnetic field, $c$ the speed of light, and $e$ and $m_i$ the proton charge and mass; 
in this limit, the wavenumber of the fastest growing mode is $k_\mathrm{max} d_i \le 1$ and reads
\begin{equation}
k_\mathrm{max} d_i = \frac{1}{2} \frac{n_\mathrm{CR}}{n_0} \frac{v_d}{v_A} \equiv \frac{1}{2}\tilde{J}_\mathrm{CR}. 
\label{eq:kMaxBell}
\end{equation}
Here $v_A \equiv B_0/\sqrt{4 \pi m_i n_0}$ is the Alfv\'en speed, $\jcr$ is the CR current normalized to $e n_0 v_A$, $n_\mathrm{CR}/n_0$ is the CR density relative to the density of the thermal ions, $v_d$ is the drift of the CR ions in the thermal ion rest frame, and  $d_i=v_A/\omega_{ci}$ is the ion inertial length. 
The corresponding growth rate in this regime is $\gamma_\mathrm{max}=v_A k_\mathrm{max}$, so that the usual Bell limit requires $\jcr\lesssim 2$.

In this paper we are interested in the high-$\jcr$ continuation of the Bell instability \citep{weidl+19a, weidl+19b}, which may be seen as a form of two-stream instability \citep{Gary1991, davidson+72,matsukiyo+03}. 
A quick inspection of Equation~\ref{eq:kMaxBell} suggests that increasing the current past $\jcr\sim 2$ would move the fastest growing mode to scales smaller than $d_i$, and with a growth rate that exceeds $\omega_{ci}$. 
This also suggests that in the high CR current regime sub-$d_i$ and electron physics may become important. 
While a significant amount of work has investigated the properties of the Bell instability \citep[e.g.,][]{bell04, reville+08a, zirakashvili+08,riquelme+09, amato+09, gargate+10, reville+13,gupta+21, zacharegkas+24}, far less was dedicated to the high-$\jcr$ regime \citep{niemiec+08,riquelme+09, weidl+19a,weidl+19b}. 
The high-$\jcr$regime is critical for describing the growth of the magnetic field immediately upstream of space and astrophysical supercritical shocks with high Alfv\'en mach number, $M_A = v_{sh}/v_A$, where $v_{sh}$ is the velocity of the incoming plasma \citep[e.g.,][]{matsukiyo+03, scholer+03, matsukiyo+06}. 
In the shock foot, in fact, about $25\%$ of the incoming particles are periodically reflected and stream along the magnetic field with a speed $\sim 2v_{sh}$ relative to the incoming plasma;
hence, even moderately large $M_A\gtrsim 10$ would naturally produce beams with $\jcr>2$ \citep[e.g.,][and references therein]{schwartz+83, caprioli+15}.

In the high-CR current regime, the shear Alfv\'en mode of the Bell instability splits into two separate ion modes with the same polarization (right-handed, circularly polarized) and helicity \citep{weidl+19a}. 
One is driven at length scales above the ion skin depth ($k d_i < 1$) and the other at length scales below it ($k d_i > 1$). 
For clarity, following the example of \cite{weidl+19a}, these modes will hereafter be referred to as the \textit{shear-Alfv\'en} and the \textit{cyclotron} modes, respectively. 
The growth rate of each of the ion modes is the same and in the limit of high-$\jcr$ is bound by the cyclotron frequency. 
Given that one of the modes driven by the high-$\jcr$ continuation of the Bell instability does fall below the ion skin depth, a kinetic treatment of the instability is crucial to determine any nonlinear interference from thermal effects. This nonlinear interference is crucial when trying to understand the saturation of the instability.

In this work, we use fully-kinetic particle-in-cell (PIC) simulations to investigate both the nonlinear interference and the saturation to obtain a complete picture of the instability's nonlinear evolution in the high-$\jcr$ regime. 
We find that, despite the presence of an electron-scale mode in addition to the two ion modes, the ion modes dominate the evolution and the saturation of the system.
Contrary to the lower-current (Bell) regime, the instability saturates below the point where the pressure in the amplified magnetic fields becomes comparable to the anisotropic CR momentum flux, as suggested by \cite{bell04} and validated by hybrid simulations \citep{zacharegkas+24}.

This result is consistent with the lower saturation levels observed first by \cite{niemiec+08} and then discussed by \cite{riquelme+09}.
The effect has been reported also in hybrid simulations, \citep{weidl+19b, zacharegkas+19p}. 
Whether hybrid simulations can capture the correct saturation of this instability, which has fastest growing modes at electron scales, remains to be validated in a fully kinetic approach.  


Our full-PIC simulations of the high-$\jcr$ streaming instability suggest that its saturation is driven by the pressure anisotropy generated by ion cyclotron heating.
This anisotropic heating, negligible in the Bell regime, is found to drive mirror modes, which nonlinearly suppress the instability well before the CRs run out of free energy/momentum.

In Section~\ref{sec:simulationsetup} we present our computational setup, validating it on the Bell instability case. 
We then discuss how the modes in the high-$\jcr$ regimes differ from the Bell modes in Section~\ref{sec:baselineRun}. 
In Section~\ref{sec:elecEn}, the energization of electrons in the high-$\jcr$ regime is studied, followed by a discussion of how the damping driven by the dynamics of the thermal ions ultimately leads to the saturation in Section~\ref{sec:saturation}. 
We extend our results to 2D simulations in Section~\ref{sec:2D}, before concluding in Section~\ref{sec:conclusion}.

\section{Simulation Setup\label{sec:simulationsetup}}

The simulations in this paper were run using the PIC code \texttt{Tristan-MP} \citep{spitkovsky05}. 
The initial magnetic field is always along the $\mathbf{x}$ direction, 
such that $\mathbf{B} = B_0 \mathbf{x}$. 
We consider a population of CRs streaming in the $\mathbf{x}$ direction with drift velocity $\mathbf{v_d} = v_d \mathbf{x}$ relative to the thermal plasma, initially at rest. 
In their rest frame, CRs have an isotropic, monochromatic momentum $p_\mathrm{iso} = m_i \gamma_\mathrm{iso} v_\mathrm{iso}$. 
Simulations are initialized with equal ion and electron temperatures, $T_i = T_e$, and periodic boundary conditions in every dimension. 
The simulations are undriven, meaning that relaxation occurs without any additional input of CRs. 

The time step is set by the speed of light and grid spacing, such that $\Delta t = 0.04~\omega_{pe}^{-1}$, where $\omega_{pe} = \sqrt{4 \pi n_e e^2/m_e}$ is the electron plasma frequency, with $m_e$ denoting the electron mass.  For all runs, the grid spacing was chosen such that $\Delta x = 0.2~d_e$, where $d_e = c/\omega_{pe}$ is the electron skin depth.
 
All ions mentioned in this paper are assumed to be protons and the magnetization was chosen such that the plasma $\beta =(8 \pi n_e k_\chg{\mathrm{B}} T_i)/B^2 = 1.28$\chg{, where $k_\mathrm{B}$ is the Boltzmann constant.} 
Both 1D and 2D simulations are included in this paper, with the 1D simulation size $L$, encompassing many unstable modes ($L k_\mathrm{max} \in [28, 1399]$ using the prediction of the fastest growing mode from Equation~\ref{eq:kMaxBell}), which ameliorates periodicity effects and increases the spectral resolution. 
We also ran different sizes in 2D, achieving convergence with a $600~d_e$-square box, for the chosen parameters.

Convergence tests in reduced mass ratio ($m_i/m_e$), number of particles per cell (ppc), and cells per $d_e$ were also performed. 
Based on these results, 1D runs were performed with 400 total particles per cell (ppc) and the 2D ones with 100 ppc, evenly distributed between ions and electrons. 
In order to maintain charge neutrality, but ensure sufficient statistics of the CR macro-particles, we used the same number of CR and thermal ion macro-particles, but with different weights such that, when integrated over all velocities, $n_{CR}/n_0$ matches the prescribed value. 

The runs are parameterized by the CR current, $\jcr$, as well as the ratio of the anisotropic momentum flux in CR and the magnetic pressure, $\xi$, which is defined as
\begin{equation}
    \xi = 2 \gamma_\mathrm{iso} \gamma_\mathrm{bst} \frac{n_\mathrm{CR}}{n_0} \frac{v_\mathrm{bst}^2}{v_{A,0}^2}  \left(1 + \frac{1}{3} \frac{v_\mathrm{iso}^2}{c^2}\right), 
    \label{eq:xiNewest}
\end{equation}
where  $v_\mathrm{bst}$ is the relative velocity between the CR rest frame and the rest frame of the background plasma \citep{zacharegkas+24}. 
The drift velocity of the CRs in the thermal ion's rest frame is defined with respect to $v_{bst}$ by the expression \citep{gupta+21}
\begin{equation}
    v_{d} = \frac{1}{2} \int_{-1}^{1}d \mu\frac{\mu v_{iso}+ v_{bst}}{1 + \mu v_{iso}v_{bst}/c} . 
\end{equation}
In the above equation, $p_\mathrm{iso}$ is the isotropic, monochromatic momentum in the CRs' rest frame and $p_d = m_i \gamma_d v_d$ is the momentum corresponding to the drift velocity of the CRs relative to the thermal ions. The parameters of all of the runs shown in this paper are detailed in Table~\ref{tab:my_label1} and \ref{tab:my_label2}. The runs are separated into two tables as different parameters are varied in each of the two groups. 

\begin{table}
    \centering
    \begin{tabular}{||c|c|c|c|c||}
     \hline
        $\tilde{J}_\mathrm{CR}$ & $\xi~(\times 10^4)$ & $n_\mathrm{CR}/n_0$ & $p_d/(m_i c^2)$ & $L_x \times L_y~[d_e]$\\ [0.5ex] 
        \hline\hline
        0.2  & 0.4712 & 0.00202 & 100 & $9000 \times 1$ \\
        \hline
        1.0 & 1.2159 & 0.01003&  33.3& $6000 \times 1$ \\
        \hline
        2.0  & 1.3441 & 0.02008 & 17.1 & $6000 \times 1$\\
        \hline
        4.0 & 1.3377 & 0.04035 & 8.32 & $6000 \times 1$\\
        \hline
        8.0 & 1.3158 & 0.08251 & 4.05 & $6000 \times 1$\\
        \hline
        10.0 & 1.2994 & 0.10496 & 3.18 & $6000 \times 1$\\
        \hline
        16.0 & 0.9652 & 0.18264 & 1.50 & $6000 \times 1$\\
        \hline
        18.0 & 1.1821 & 0.21418 & 1.56 & $6000 \times 1$\\
        \hline
        20.0 & 1.1366 & 0.24962 & 1.34 & $6000 \times 1$\\
        \hline \hline 
        4.0 & 1.3377 & 0.04035 & 8.32 & $600 \times 600$ \\
        \hline
        7.0 & 1.3275 & 0.07155 & 4.69 & $600 \times 600$\\
        \hline
        10.0 & 1.2994 & 0.10496 & 3.18 & $600 \times 600$\\
        \hline
        16.0 & 0.9652 & 0.18264 & 1.50 & $600 \times 600$\\
        \hline
    \end{tabular}
    \caption{Parameters for the runs in Figure~\ref{fig:BaselineBellFig1}, \ref{fig:BaselineTSIFig2}, \ref{fig:Fig3_1DScan}, \ref{fig:Fig4_ElecEn_SiddComp}(a), \ref{fig:Fig5_BSat}, and \ref{fig:Fig6_2DJcrScan}.  For all of these runs $p_\mathrm{iso}/(m_i c) = 1$, $m_i/m_e = 1836$, $\sigma = (\omega_{ce}/\omega_{pe})^2 = 0.1836$ (where $v_A/c = \sqrt{\sigma/(m_i/m_e)}$), and $\delta \gamma \equiv (v_{th, i, 0}/c)^2 = k_B T_i/(m_i c^2) = 6.4\times 10^{-5}$.}
    \label{tab:my_label1}
\end{table}

\begin{table}
    \centering
    \begin{tabular}{|c ||c|c|c|c|c|c|c||}
     \hline
        ~ &$\xi$ & $\tilde{n}_\mathrm{CR}$ & $p_d/(m_i c^2)$ & $\tilde{p}_\mathrm{iso}$ & $\delta \gamma$ & $m_\mathrm{r}$\\ 
        ~ &~ & ~ & ~ & $\times10^{-3}$ & $\times10^{-5}$ & ~\\ 
        [0.5ex] 
        \hline\hline
        A &393 & 1E-1 & 0.416 & 400 & 6.4 & 100\\
        \hline
        B &349 & 1E-2&  0.436 & 4.5 & 6.4 & 1836\\
        \hline
        C &393 & 1E-1 & 0.416 & 400 &6.4 & 1836\\
        \hline
        D& 1,236 & 2.6E-2 & 0.381 & 4.5& 36 & 1836\\
        \hline 
    \end{tabular}
    \caption{Parameters for the runs in Figure~\ref{fig:Fig4_ElecEn_SiddComp}(b) and (c). For all of these runs $\tilde{J}_\mathrm{CR} = 3.8$, $L_x \times L_y = [6000 \times 1]~d_e$ and $\sigma = (\omega_{ce}/\omega_{pe})^2$ such that $\beta = 2 (\delta \gamma)(m_i/m_e) / \sigma = 1.28$. The parameter $\tilde{p}_\mathrm{iso} = p_\mathrm{iso}/(m_i c)$, $\tilde{n}_\mathrm{CR} = n_\mathrm{CR}/n_0$, and $m_\mathrm{r} = m_i/m_e$. The runs in this set correspond (from the top down) to the red, green, blue, and black runs in Figure~\ref{fig:Fig4_ElecEn_SiddComp}(b).}
    \label{tab:my_label2}
\end{table}

\subsection{The Bell Instability Regime\label{sec:Bell}}

\begin{figure}[ht!]
\includegraphics[width=\linewidth]{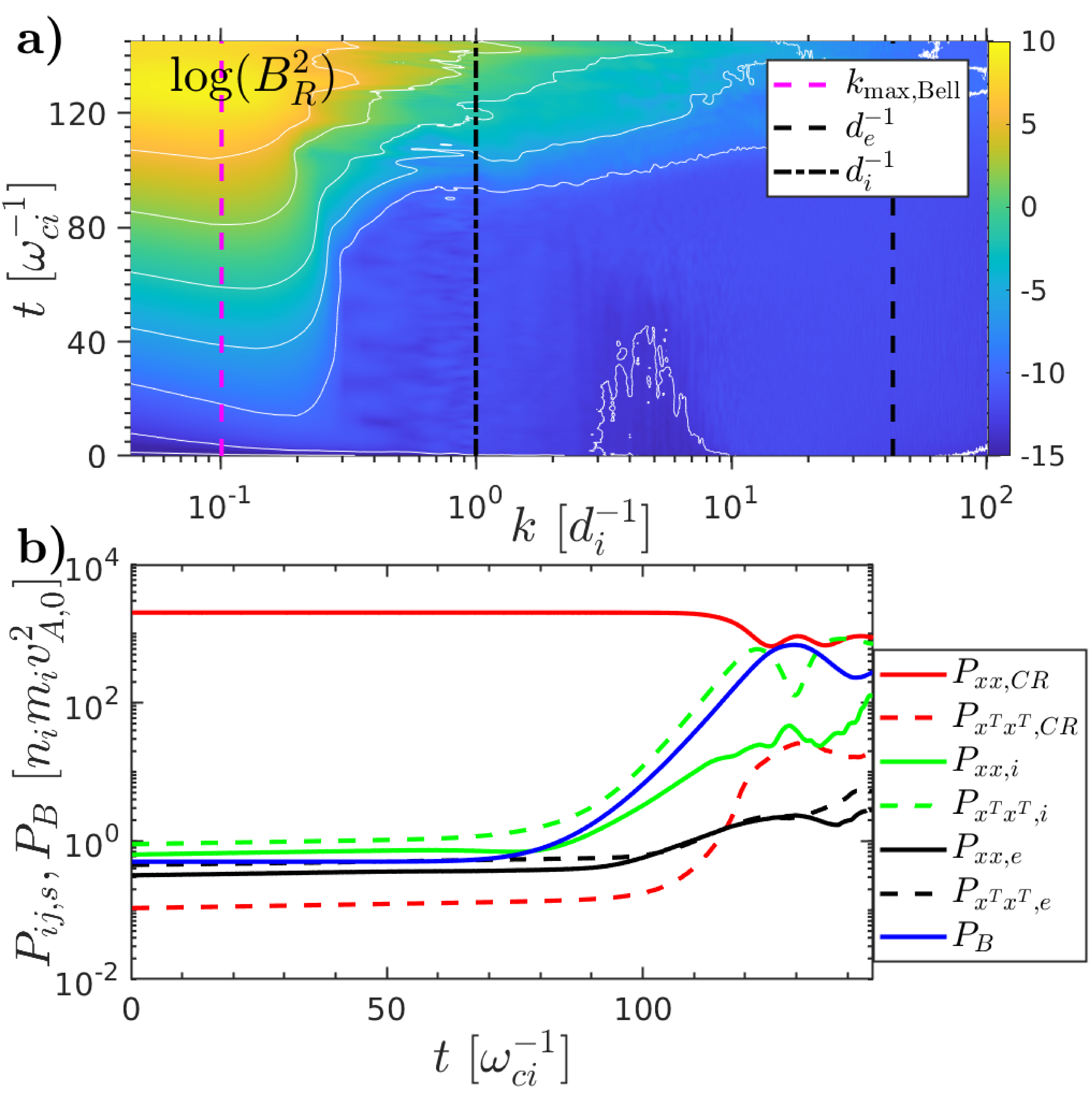}
\caption{\textbf{a)} Right-handed magnetic field magnitude (where $B_R = |\mathrm{FT}(B_y/B_0) + i\mathrm{FT}(B_z/B_0)|$ and $\mathrm{FT}(...)$ is the Fourier transform along x) as a function of time, $t$, and wavenumber, $k d_i$, for a 1D simulation of the Bell instability ($\jcr = 0.2$\chg{, $\xi = 0.47 \times 10^{4}$}). 
The dashed pink line indicates the predicted fastest growing mode (Equation~\ref{eq:kMaxBell}).
\textbf{b)} Pressure for each species. The pressure ($P_{jj, s}$) for each direction $j = x, x^T$ is calculated in the thermal ion rest frame.
The pressure in the $x$ direction and the pressure transverse to the $x$ direction, $P_{x^T x^T,s} = 0.5(P_{yy, s} + P_{zz,s})$ are both plotted along with the magnetic pressure ($P_B$). Note how at saturation $P_B$ becomes comparable to the anisotropic, driving CR pressure \citep{zacharegkas+24}.} 
\label{fig:BaselineBellFig1}
\end{figure}

An illustration of the instability's behavior in the Bell regime can be seen in Figure~\ref{fig:BaselineBellFig1}. Figure~\ref{fig:BaselineBellFig1}(a) shows the time evolution of the right-handed magnetic perturbations. The fastest growing modes are consistent with the prediction of Equation~\ref{eq:kMaxBell} (pink dashed line), and there is no growth at scales below $d_i^{-1}$. 
The instability saturates when the pressure in amplified magnetic fields, CRs, and thermal plasma are of the same order of magnitude, as found in hybrid simulations \citep{zacharegkas+24}.
Given the beam-like nature of the prescribed CR distribution ($P_{yy, CR}, P_{zz, CR} \ll P_{xx, CR}$, where $P_{jj, s} = \int d^3v d^3x \gamma(v) (v_j - V_{d,thi, j})^2 f_s$ and $V_{d,thi,j}$ is the mean velocity of the thermal protons in direction $j$), this occurs when $P_B \sim P_{xx, CR}$, as can be seen in Figure~\ref{fig:BaselineBellFig1}(b).  
In terms of the free energy in the CRs, this means that $\delta B/B_0 \sim \xi^{1/2}/2$ \citep{zacharegkas+24}. 

\section{Modes present in the high-$J_\mathrm{CR}$ regime\label{sec:baselineRun}}

\begin{figure}[ht!]
\includegraphics[width=\linewidth]{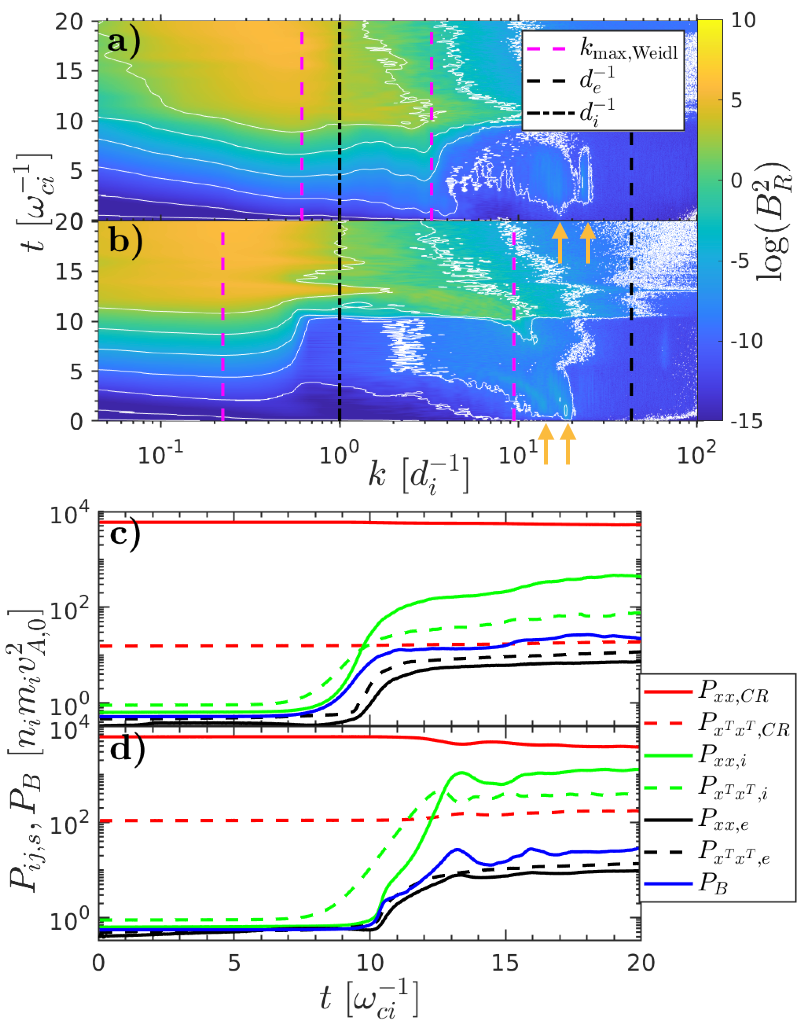}
\caption{Right-handed magnetic field magnitude as in Figure~\ref{fig:BaselineBellFig1} for 1D simulations where \textbf{(a)} $\tilde{J}_\mathrm{CR} = 4.0$ and \textbf{(b)} $\tilde{J}_\mathrm{CR} = 10.0$. 
\chg{For these simulations, $\xi = 1.34 \times 10^4$ and $1.30 \times 10^4$ respectively.}
The analytical prediction for the fastest growing modes in the high-$\jcr$ regime \citep{weidl+19a}  are indicated with pink, dashed lines. 
Orange arrows denote the positions of electron modes. Pressure for each species and energy in the magnetic field is shown for \textbf{(c)} $\tilde{J}_\mathrm{CR} = 4.0$ and \textbf{(d)} $\tilde{J}_\mathrm{CR} = 10.0$. 
In the high-$\jcr$ regime both the modes and the energy partition at saturation are different from the Bell regime.} 
\label{fig:BaselineTSIFig2}
\end{figure}

For $\jcr > 2$ the modes that are driven are significantly different than in the traditional Bell regime, as illustrated in the two high-$\jcr$ simulations in Figure ~\ref{fig:BaselineTSIFig2}. 
We identify four distinct modes.
Two of them grow on ion gyration timescales, which we refer to as \textit{ion modes};
indicated by magenta lines in Figures 2(a) and (b), they correspond to the high-$\jcr$ continuation of the Bell instability, as pointed out by  \cite{weidl+19a}. 
The other two modes grow much faster than the ion modes and persist only for a short duration; 
since they appear at length scales much smaller than $d_i$, we refer to them as \textit{electron modes} (orange arrows). 

While the electron modes ultimately contain only a negligible fraction of the total magnetic power, for $t \leq \omega_{ci}^{-1}$ they grow faster than the ion modes. 
Note that in the simulation with $\tilde{J}_{CR} = 10.0$ (Figure~\ref{fig:BaselineTSIFig2}(b)) only one ion mode is noticeable, the reason being that the growth of the ion cyclotron mode with $k d_i > 1$ is hindered by the saturation of the much faster electron modes that are driven at comparable wavenumbers. 


To better understand the nature of the modes and their connection with Bell's modes, it is useful to consider the dispersion relation for a cold thermal plasma with a drifting ``beam"/CR population.
Following the derivation in \cite{weidl+19a} (see their equation 19), we assume that the modes have a frequency $\omega = \omega_R + i \gamma$ such that $|\omega| \ll \omega_{ce}$ and obtain
\begin{align}
(\omega^2 &- k_{\parallel}^2 c^2) \frac{n_e}{\omega_p^2} \nonumber\\
&= n_0 \frac{\omega}{\omega + \omega_{ci}} - n_e \frac{\omega - V_e k_{||}}{\omega_{ci}} + n_\mathrm{CR} \frac{\omega - v_d k_{\parallel}}{\omega - v_d k_{\parallel} + \omega_{ci}} ,
\label{eq:Weidl19}
\end{align}
where $\omega_p = (n_e e^2/m_i)^{1/2}$ and $V_e = n_\mathrm{CR} v_d/(n_0 + n_\mathrm{CR})$. 

To recover the Bell instability for the low-$\jcr$ regime, two assumptions are made. 
First, modes are assumed to be in the MHD frequency range, i.e., $|\omega| \ll \omega_{ci}$, which approximates the contribution of the thermal ions (first term on the right-hand side of Equation~\ref{eq:Weidl19}) as $\omega_{pi}^2(\omega/\omega_{ci} - \omega^2/\omega_{ci}^2 + \mathcal{O}(\omega^3/\omega_{ci}^3))$. 
The second assumption is that the frequency of the most unstable mode is far from gyroresonance with the beam/CR ions, i.e.,  $|\omega - k_{||}v_d| \ll \omega_{ci}$. This approximates the contribution of the beam/CR ions (the last term on the right-hand side of Equation~\ref{eq:Weidl19}) as zero. 

\begin{figure}[ht!]
\includegraphics[width=\linewidth]{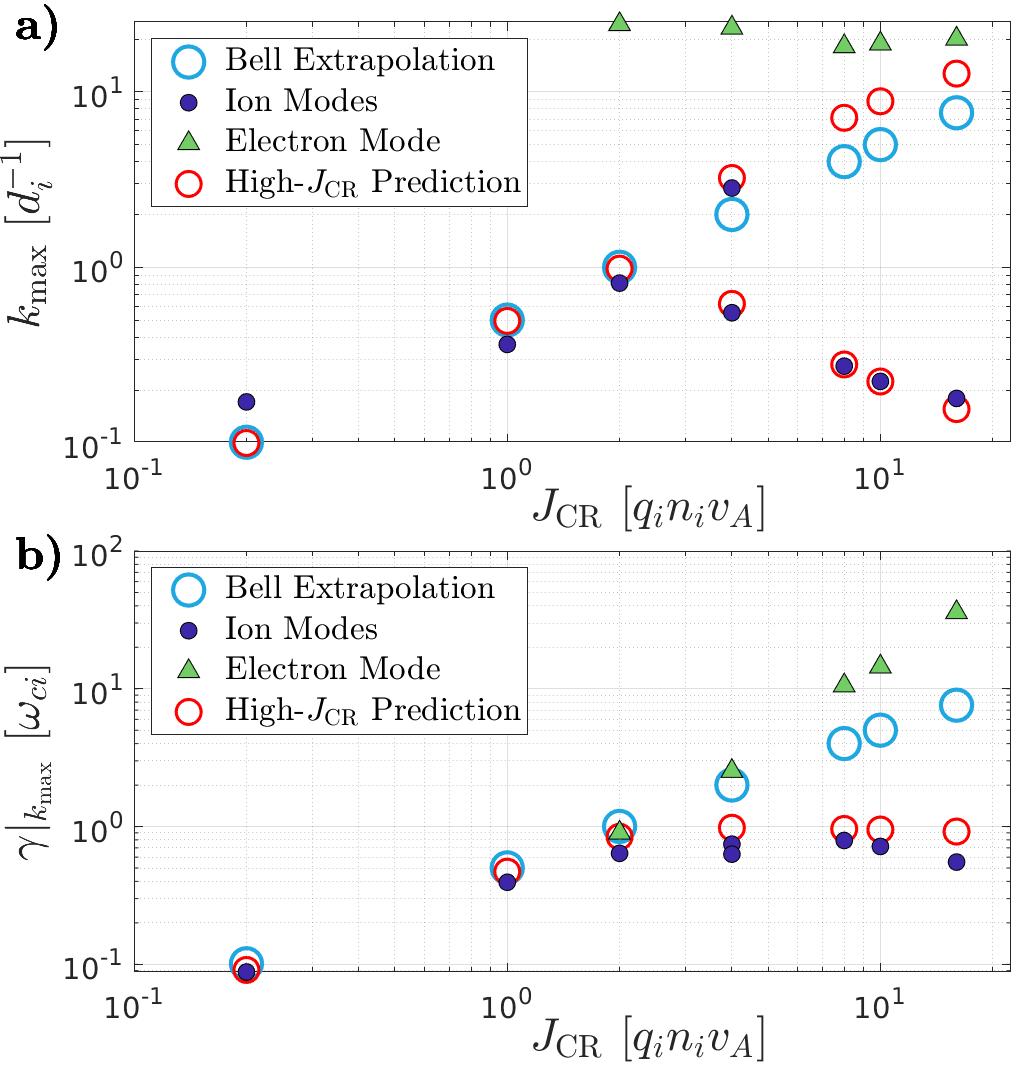}
\caption{\textbf{a)} $k_\mathrm{max} d_i$ for both ion-scales (dark blue circles) and electron-scales (green triangles) with the extrapolation of the Bell prediction (Equation~\ref{eq:kMaxBell}) to the high-$\jcr$ regime (open cyan circles) and with the predictions for the more general expression in Equation~\ref{eq:Weidl19} (open red circles), as a function of $\tilde{J}_\mathrm{CR}$. \chg{Here $\xi$ is held relatively constant, $\mathcal{O}(\xi) \sim 10^4$.}
$k_\mathrm{max}$  is averaged over $t < \omega_{ci}^{-1}$ for the electron-scale mode and $t \in (\omega_{ci}^{-1}, t_{\rm saturation})$ for the ion-scale mode. 
For $\jcr > 4$ only one of the ion modes could be identified, because of the nonlinear interference with the electron modes. 
\textbf{b)} As in the panel above, but for the growth rate of both ion and electron modes; the ion modes match well with the prediction by \cite{weidl+19a}.} 
\label{fig:Fig3_1DScan}
\end{figure}

However, as shown by \cite{weidl+19a}, in the high-current regime neither approximation holds, since both the real and imaginary components of the frequency are of the same order as the ion cyclotron frequency.
We solve Equation~\ref{eq:Weidl19} numerically and compare the predicted wavenumber and growth rate of the fastest-growing modes to the results of 1D simulations performed over a range of $\jcr$ (see Figure~\ref{fig:Fig3_1DScan}).
Good agreement is found between predictions (open red circles) and the ion modes observed in simulations (filled blue circles), both in wavenumber and growth rate of the fastest growing mode.
For computational simplicity, when estimating $k_\mathrm{max}$ and $\gamma|_\mathrm{k_{max}}$ from the simulations, we consider \chg{the modes which contain the most power at a given point in time ($t_0$) for a given range of $k\in k_\mathrm{Range}$, \textit{i.e.} $k|_{\max P(t_0, k \in k_\mathrm{Range})}$, where $k_\mathrm{Range}$ varies depending on the mode. For example, for $k_\mathrm{max}$ of the shear-Alfv\'en mode at a given point in time only wavenumbers in the range $k_\mathrm{Range} = [10^{-2}, 10^0] d_i^{-1}$ were considered.} 

\chg{

We note that while $p_{iso}/(m_i c^2) = 1$ is held constant for every simulation in Table~\ref{tab:my_label1}, in order to vary $\jcr$ as a function of relatively constant $\xi$ $p_d/(m_i c^2)$ varies from 100 to 1.3. 
As such, even though the analytic prediction in Equation~\ref{eq:Weidl19} for the wavenumber, growth rate, and frequency of the ion modes was derived assuming a cold plasma ($p_{iso} \ll p_d$) these simulations show that this analytic prediction is a good descriptor of the plasma even where $p_{iso} \sim p_d$.}

The wavenumber of the electron mode is constant as a function of the CR current, but its growth rate is directly proportional to the CR current. 
Both electron modes are identified in Figure~\ref{fig:BaselineTSIFig2}(a) and (b) by the orange arrows, but only the electron mode with the most power is included in Figure~\ref{fig:Fig3_1DScan} for clarity. 
Note that electron modes depend on retaining a finite electron temperature \cite{matsukiyo+03, bret09, gupta+24a}, so they are not accounted for in Equation~\ref{eq:Weidl19}.
Fully characterizing them is beyond the scope of this paper, especially because they are found not to be important for the overall saturation.

\section{Electron Energization\label{sec:elecEn}}

\begin{figure}[ht!]
\includegraphics[width=\linewidth]{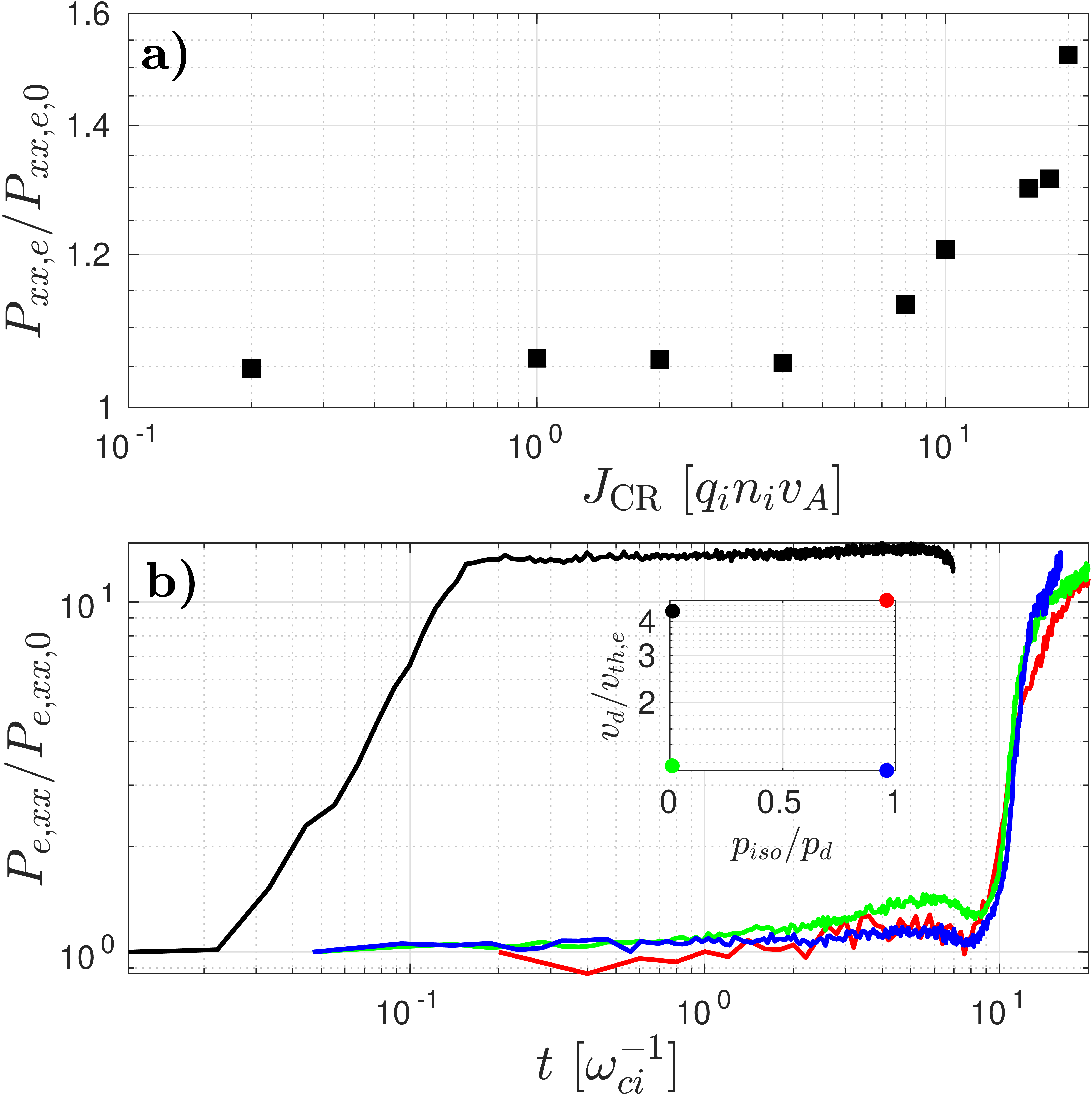}
\caption{\textbf{a)} Parallel electron pressure increase due to the electron-scale mode for $t < \omega_{ci}^{-1}$ as a function of CR current.  
\textbf{b)} Parallel electron pressure for four simulations with $\tilde{J}_\mathrm{CR} = 3.8$, $\xi \in [349, 1236]$, and $\beta = 1.28$, but with varying values of $v_d/v_{th, e}$ and $p_{iso}/p_d$ (as in the inset figure). 
Note how rapid (i.e., $\lesssim \omega_{ci}^{-1}$) electron heating requires both  $v_d/v_{th,e} \gg 1$ and $p_{iso}/p_d \ll 1$ \citep{gupta+24a}.}
\label{fig:Fig4_ElecEn_SiddComp}
\end{figure}

For the parameters considered above, the electron modes contribute relatively little to the electron energization.
Over the first $\omega_{ci}\chg{^{-1}}$, electron heating is roughly isotropic and relatively small, as shown in Figure~\ref{fig:Fig4_ElecEn_SiddComp}(a).  
Despite the increase in the amount of heating with increasing CR current, even the strongest response corresponds to only $\sim 50\%$ increase in the initial parallel electron pressure, orders of magnitude below the energy in the CRs.  

As can be seen in Figure~\ref{fig:BaselineTSIFig2}(b) and (d), most of the electron heating occurs several cyclotron times after the electron mode has saturated, on the timescales of the saturation of the ion-scale mode. 
We conclude that, while in the high-$\jcr$ limit the electron modes are faster, it is the ion response that dominates both particle heating and magnetic field dynamics after the first $\omega_{ci}^{-1}$. 

There is a special case in the high-CR current regime where the electron dynamics may play an important role in the system, though. 
When both the electrons are very cold relative to the drift velocity of the CRs, $v_d/v_{th,e} \gg 1$ and the CRs themselves are extremely cold relative to their own drift velocity, $p_{iso}/p_d \ll 1$, an electrostatic instability is generated that causes substantial electron heating. 
This effect is discussed in the context of the shock foot, e.g., by \cite{matsukiyo+03,scholer+03,gupta+24a} and demonstrated in Figure~\ref{fig:Fig4_ElecEn_SiddComp}(b). 
The behavior of the parallel electron pressure for four simulations with the same CR current, $\jcr = 3.8$, and with $\xi \in [349, 1236]$ are plotted in all combination of the limits of $v_d/v_{th,e}$ and $p_{iso}/p_d$, with the run parameters detailed in Table~\ref{tab:my_label2}. For runs A (red, $v_d/v_{th,e} \gg 1$ \& $p_{iso}/p_d \sim 1$), B (green, $v_d/v_{th,e} \ll 1$ \& $p_{iso}/p_d \ll 1$), and C (blue, $v_d/v_{th,e} \ll 1$ \& $p_{iso}/p_d \sim 1$), there is negligible electron heating until a sharp jump when the ion mode saturates, occurring around $10~\omega_{ci}^{-1}$. 
For run D (black, $v_d/v_{th,e} \gg 1$ \& $p_{iso}/p_d \ll 1$) there is an order of magnitude increase in the electron heating, which saturates on the same timescale as the electron mode ($t \omega_{ci} < 1 $). 
This may be important for electron heating in the shock foot and crucial for electron injection \citep[e.g.,][and references therein]{hoshino+02,scholer+03,bret+10, muschietti+17,bohdan+20a, gupta+24a}.


\section{Magnetic Field Saturation\label{sec:saturation}}

\begin{figure}[ht!]
\includegraphics[width=\linewidth]{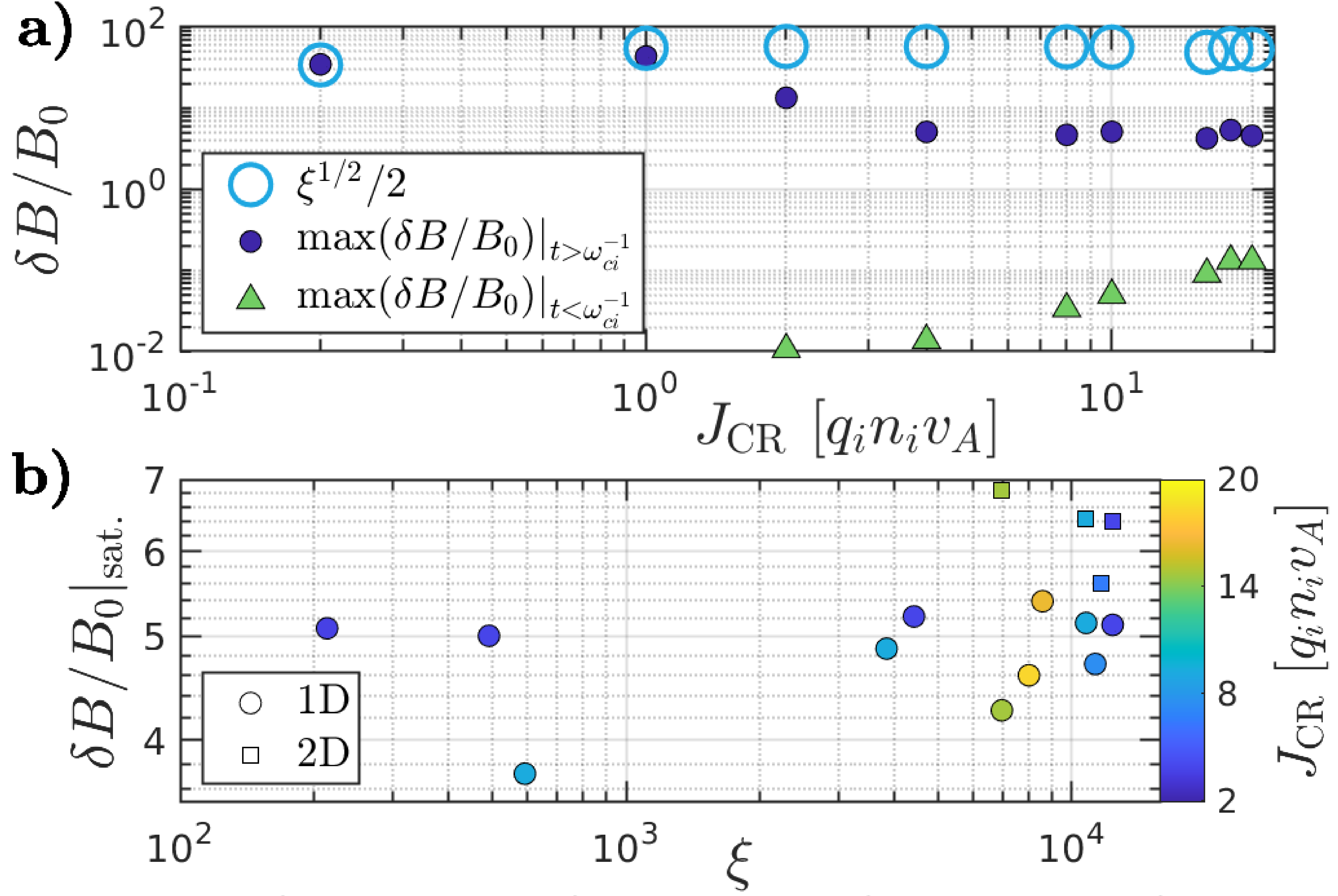}
\caption{\textbf{(a)} Maximum of the transverse magnetic field amplitude, $\delta B/B_0$, for electron mode time-scales ($t < \omega_{ci}^{-1}$, green triangles) and ion mode time-scales ($t > \omega_{ci}^{-1}$, dark blue circles), as a function of CR current, $\tilde{J}_\mathrm{CR}$. The expectation for the Bell instability, $\delta B/B_0 = \sqrt{\xi}/2$ \citep{zacharegkas+24}, is plotted with cyan open circles.
\textbf{(b)} Transverse magnetic field amplitude at saturation as a function of $\xi$ for $\tilde{J}_\mathrm{CR} > 2$. 
Across a wide range of $\xi$ and $\jcr$, the high-$\jcr$ runs saturate at a roughly constant value, about one order of magnitude below the extrapolation of the Bell prediction. } \label{fig:Fig5_BSat}
\end{figure}

A notable result of our simulations is that, unlike in the Bell regime, in the high-$\jcr$ limit the same amount of initial CR free energy (same $\xi$), leads to smaller magnetic fields at saturation, as shown in Figure~\ref{fig:Fig5_BSat}(a).
Though this has been reported before \citep[e.g.,][]{niemiec+08, zacharegkas+19p}, this is the first time that the final magnetic field is contrasted with the physical picture that the saturation ensues only when the magnetic pressure achieves equipartition with the initial anisotropic CR momentum flux, as detailed in \cite{zacharegkas+24}.

\begin{figure}[ht!]
\includegraphics[width=\linewidth]{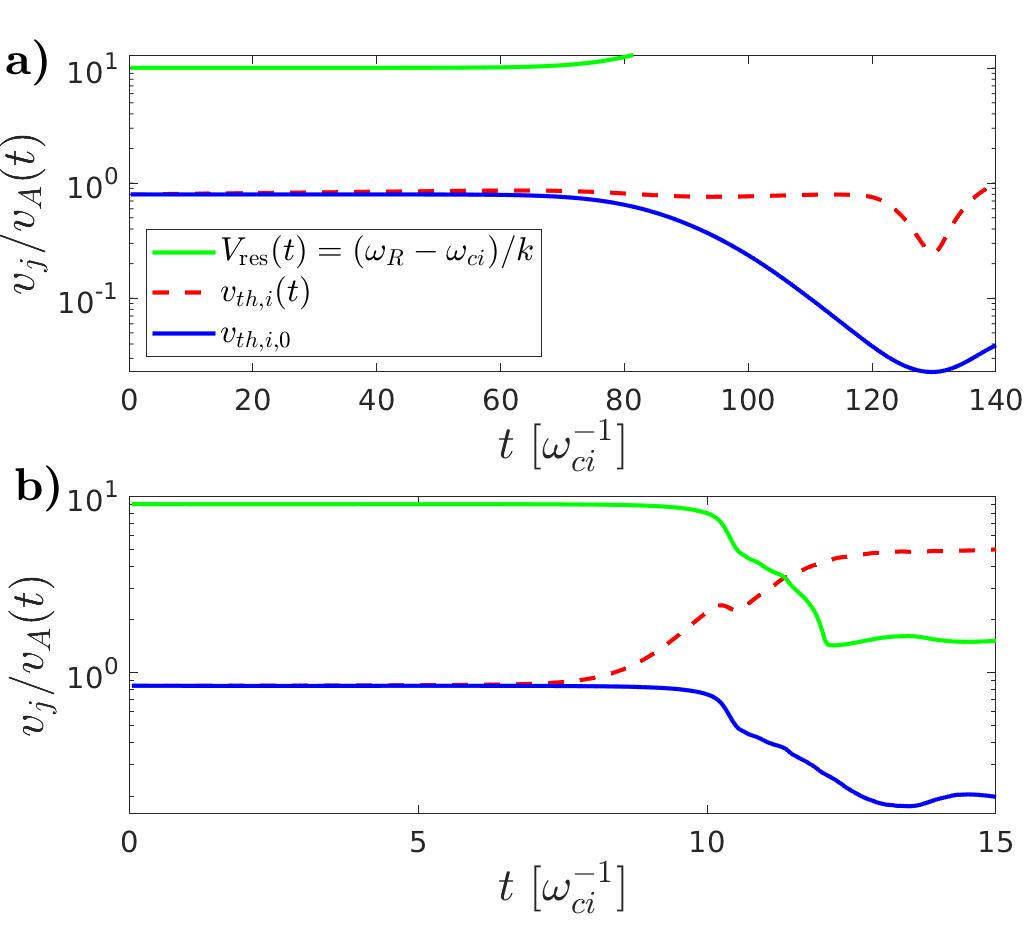}
\caption{\textbf{a)} Time-dependent thermal velocity (red, dashed), resonant velocity (green), and initial thermal velocity (blue) as normalized by the time-dependent $v_A$ for \chg{$\jcr = 0.2$} \chg{($\xi = 0.47 \times 10^4$)}. \textbf{b)} Same plot but for $\jcr = 4.0$ \chg{($\xi = 1.34 \times 10^4$)}. Both the runs in the Bell and high-$\jcr$ regime see non-adiabatic heating (red $>$ blue). With time the velocity at which the particles are resonant with the mode with the most power (green) moves closer to the center of the distribution for the high-$\jcr$ regime, allowing more particles to experience ion cyclotron damping. This also corresponds to an increase in the plasma beta (red) not seen in the Bell regime.} \label{fig:VthVg}
\end{figure}

\begin{figure}[ht!]
\includegraphics[width=\linewidth]{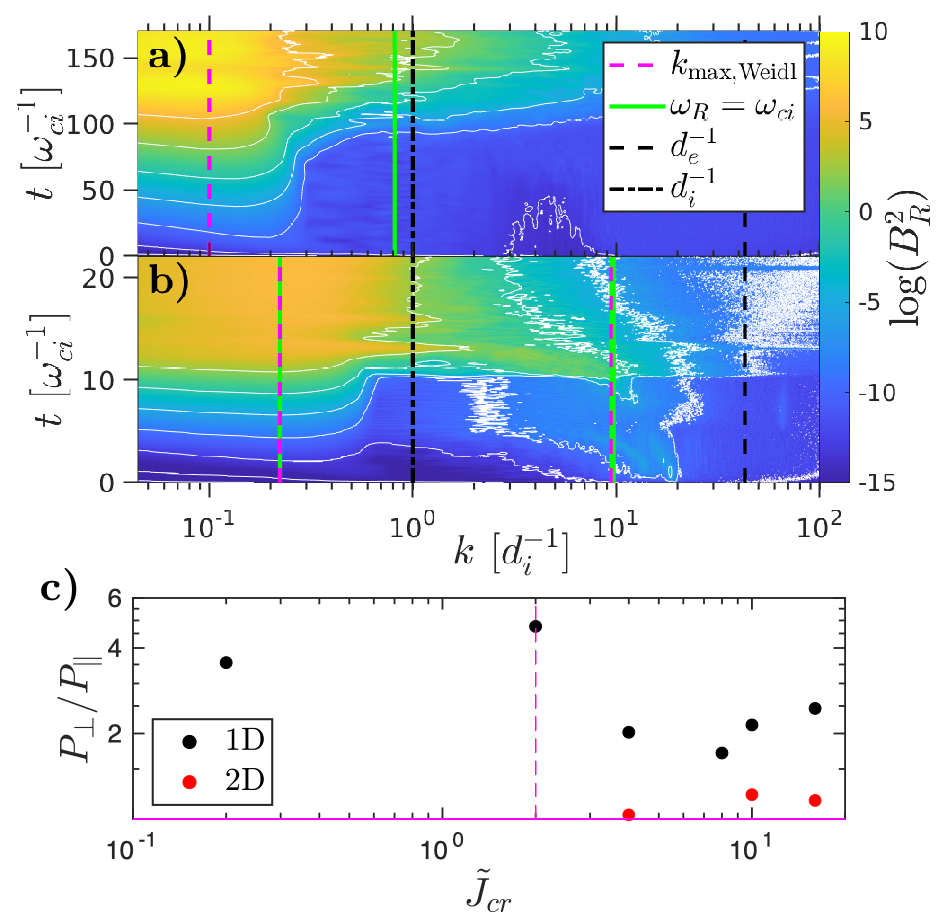}
\caption{\textbf{(a)} Right-handed magnetic field as a function of wavenumber and time for $\tilde{J}_\mathrm{CR} = 0.2$ \chg{($\xi = 0.47 \times 10^4$)}. The green line is the analytic prediction from Equation~\ref{eq:Weidl19} for where $\omega_\mathrm{R} = \omega_{ci}$, where ion cyclotron damping should be strongest, using the initial simulation parameters. The pink, dashed line is where the mode with the most power is predicted by Equation~\ref{eq:Weidl19} for the initial simulation parameters. \textbf{(b)} The same quantities as above for a simulation with $\tilde{J}_\mathrm{CR} = 10$ \chg{($\xi = 1.30 \times 10^4$)}. For the high-$\jcr$ case ion cyclotron heating is at a maximum where most of the mode power is driven. This is not true for the case in the Bell regime. \textbf{(c)} Pressure anisotropy of each simulation at saturation as a function of $\jcr$\chg{, for $\mathcal{O}(\xi) \sim 10^4$}. The horizontal magenta line is plotted at isotropy ($P_{\perp}/P_{\parallel} =1$) and the vertical, dashed magenta line denotes the boundary between the Bell limit ($\jcr < 2$) and the high-$\jcr$ limit ($\jcr > 2$). } \label{fig:Fig6_BCycDamp}
\end{figure}

Notably, across a large range of input parameters spanning one order of magnitude in $\jcr$ and more than two in $\xi$, the magnetic field saturates at the same level, roughly, $\delta B/B_0 \sim 5$. 
This can be seen in Figure~\ref{fig:Fig5_BSat}(b), where the transverse magnetic field at saturation for all the runs in Table~\ref{tab:my_label1} and \ref{tab:my_label2} with $\jcr \ge 2$ that are not in the electrostatic instability regime (\textit{i.e.,} have both $v_d/v_{th,e} \gg 1$ \& $p_{iso}/p_d \ll 1$) are plotted. 

The saturation at this point cannot be accounted for by any linear modification or continuation of the ion mode. 
Past work has successfully shown that including thermal effects from including a finite temperature can affect the growth rate of the Bell instability \citep[see, e.g.,][]{reville+08a,everett+11, marret+21}. 
To test whether this is at play in this case, we utilized the \texttt{PLUME} framework, a linear solver that calculates the unstable modes driven by an arbitrary number of bi-Maxwellian distributions \citep{Klein2015}. 
Despite being non-relativistic, \texttt{PLUME} was able to reproduce the initial growth rates and wavenumbers for the ion modes that were both measured and predicted by \cite{weidl+19a} for relativistic beams in the cold plasma limit. This is likely because for this instability both the growth rate and the wavenumber of the fastest growing mode depend only on $\jcr$.  

Plugging into \texttt{PLUME} the temperatures, magnetic field, and relative drift measured in the simulations at saturation, the expected growth drops from $\sim 1.0~\omega_{ci}$ to $\sim 0.1~\omega_{ci}$, due to the combination of the decreased relative drift and the finite-temperature effects. 
However, in simulations we do not observe further growth on this timescale.
Therefore linear theory (even including finite temperature corrections) does not predict that the system will be at marginal stability.
Yet, our simulations show that the plasma is not able to tap into the remaining free energy in the CRs.

\begin{figure}[ht!]
\includegraphics[width=\linewidth]{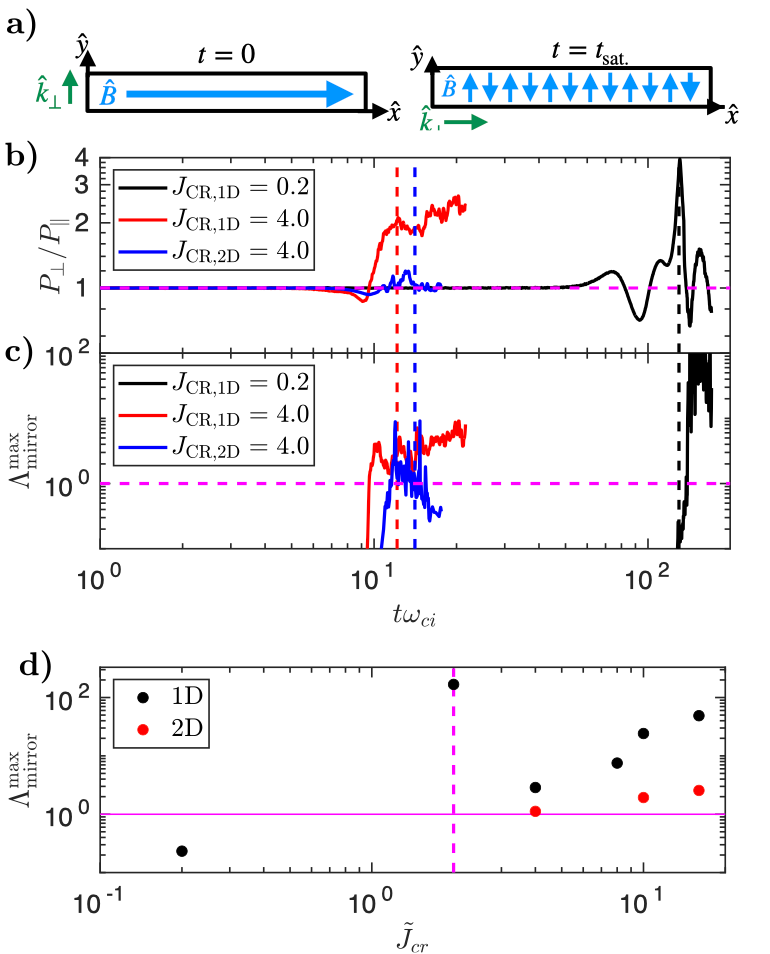}
\caption{\textbf{a)} Cartoon of the $B$-field reorientation from the initial parallel configuration (left) to the one at saturation (right), which allows the development of the mirror modes. 
\textbf{b)} Pressure anisotropy in thermal protons, $P_{\perp}/P_{\parallel} = 0.5(P_{\perp,1} + P_{\perp,2})/P_{\parallel}$ averaged over the entire box as a function of time, where $P_{\perp,1}$ and $P_{\perp,2}$ are the pressure in the two directions perpendicular to the local $\mathbf{B}$. 
Dashed vertical lines correspond to the saturation time of each simulation. The dashed pink horizontal line corresponds to isotropy. 
\textbf{c)} Maximum value of $\Lambda_\mathrm{mirror}(t)  = \beta_{\perp,i}(t)(P_{\perp, i}(t)/P_{\parallel, i}(t) - 1)$ in the simulation domain as a function of time. 
For computational efficiency only the thermal ion contribution is included - the contribution of the thermal electrons is negligible for all cases in this paper. 
Values above the threshold $\Lambda_\mathrm{mirror}^\mathrm{max} = 1$ (pink solid line) correspond to a plasma that is, at least locally, prone to the mirror instability. 
Note that despite the presence of pressure anisotropy in both cases, only the high-$\jcr$ runs become mirror unstable before saturation. \textbf{d)} The value of $\Lambda_\mathrm{mirror}^\mathrm{max}$ at saturation for the runs in Table~\ref{tab:my_label1}.} \label{fig:LamMirror}
\end{figure}

The key to explaining this difference in magnetic field saturation may be found in the \textit{nonlinear} response of the thermal ions. 
Retaining the finite temperature of the background ions opens up the possibility of another route of energy transfer between the thermal ions and the magnetic field, namely ion cyclotron damping, which occurs when the following resonant condition is met 
\begin{equation}
    \omega = k_{\parallel} v_{\parallel} - m \omega_{cj} \qquad \qquad (m = \pm 1, \pm 2, ...),
    \label{eq:icRes}
\end{equation}
where $\omega_{cj}$ is the cyclotron frequency for species $j$. 
The $m = 0$ case corresponds to Landau damping, where the resonance condition leads to heating parallel to the magnetic field. 
However, for $|m| > 0$ the perpendicular fluctuations of the particles and the fields become in phase with each other, resulting in net heating in the perpendicular direction \citep{Stix1992, Gary1994}. 

To characterize the velocities that are most impacted by ion cyclotron heating, we introduce an effective resonant velocity, defined as $V_\mathrm{res}(t) = (\omega_R(t) - \omega_{ci})/k(t)$\chg{. Here, $\omega_R(t)$ and $k(t)$ of the fastest growing mode are calculated for each time step by using Equation~\ref{eq:Weidl19} with the instantaneous values of $v_d(t)/v_A(t)$. Note that the total magnetic field $B(t)$ is used at each time step to calculate $v_A(t)$. This effective resonant velocity should be considered representative as in actuality there exist a band of modes, centered around the fastest growing mode, which all heat the plasma.}

Figure~\ref{fig:VthVg} shows such a resonant velocity \chg{for the fastest growing mode} (green lines) \chg{for ($m = -1$)} along with the initial \chg{($v_{th, i, 0}$)} and instantaneous \chg{($v_{th,i}(t)$)} thermal speed of the plasma (blue and \chg{dashed} red lines, respectively), all normalized by the time-dependent Alfv\'en speed, calculated in the amplified magnetic field. 
\chg{For these simulations the effective resonant velocity for the $|m|\ge2$ modes was far enough from the core of the thermal ion and electron distributions that they have a minimal effect on the damping. }
\chg{While the $m = 1$ mode exhibits a similar behavior to the $m=-1$ mode, only positive velocities are plotted for ease of interpretation.}
\chg{The $v_{th,i,0}/v_A(t) \propto 1/B(t)$ line is added to the figure to trace when saturation occurs and expresses the change in thermal velocity expected in the absence of non-adiabatic heating processes.}

While in the Bell regime the ion thermal speed grows together with the Alfv\'en speed (leading to the constant red, dashed line in Figure~\ref{fig:VthVg}(a)), in the high-$\jcr$ case (bottom panel) the ion thermal speed $v_{th}(t)/v_A(t)$  increases with time, attesting to the presence of additional heating.
However, even as the magnetic field increases, for the Bell case the effective resonant velocity moves away from the center of the distribution and the relative amount of particles participating in this heating process does not increase.
On the other hand, for the high-$\jcr$ case, as the magnetic field increases, the resonant velocity moves close to the center of the distribution, resulting in a very efficient ion heating. 
Additionally, because $\beta(t) \propto (v_{th}(t)/v_A(t))^2$, in the Bell case the plasma beta remains constant, whereas it increases in the high-$\jcr$ case.

We note that while the linear prediction used to produce the effective resonant velocity accurately predicts the modes present in the high-$\jcr$ cases through saturation, for the Bell case this linear prediction is only valid to roughly $90 \omega_{ci}^{-1}$. 
This, combined with observations in earlier works of a non-zero real frequency associated with the Bell mode \citep{riquelme+09, gargate+10}, suggests that near saturation there may be some non-zero amount of heating that is due to ion cyclotron heating as well. 
However, given the difference of behavior in $v_{th}(t)/v_A(t)$ and the saturation at equipartition, we conclude that ion cyclotron heating does not play a significant role in the evolution and saturation in the Bell regime. 

Additionally, ion cyclotron damping is maximal when the real frequency of the wave is on the order of the cyclotron frequency \citep{Gary1994,Stix1992}. 
For the Bell instability (where $\omega_\mathrm{R} \ll \omega_{ci}$) the scales at which cyclotron damping is most efficient is not near the scales where the waves are being driven, unlike in the high-$J_\mathrm{CR}$ regime, where $\omega_\mathrm{R} \sim \omega_{ci}$ and the cyclotron damping occurs on the exact same scales where most of the growth of the magnetic field is occurring. 
As such, in the high-$\jcr$ regime there are plenty of fluctuations at the cyclotron frequency to facilitate ion cyclotron damping that do not exist in the Bell case. 
This difference is illustrated in Figure~\ref{fig:Fig6_BCycDamp}(a) and (b), where the green lines denote the scales at which ion-cyclotron heating is maximal based on the linear predictions for $\omega_R = \omega_{ci}$ from Equation~\ref{eq:Weidl19}. 
In the Bell regime (Figure~\ref{fig:Fig6_BCycDamp}(a)), the scales at which the modes are growing are far larger than the scales at which damping is occurring, whereas in the high current regime example in Figure~\ref{fig:Fig6_BCycDamp}(b) the scales at which the modes are growing are exactly the same as where the damping is occurring. 

As would be expected from ion-cyclotron damping, the corresponding thermal ion heating is very anisotropic, with $P_{\perp}/P_{\parallel} \gg 1$ for all cases, both in the Bell and high-$\jcr$ regimes. This can be seen in Figure~\ref{fig:Fig6_BCycDamp}(c), where the value of the pressure anisotropy at saturation is plotted as a function of $\jcr$.

The initial 1D setup shown in the cartoon in Figure~\ref{fig:LamMirror}(a, left) does not accommodate a wavelength perpendicular to the direction of the initial \chg{mean magnetic field ($\mathbf{B} = B_0 \mathbf{x}$). However after the ion mode has grown such that $B_y, B_z \gg B_x$ the mean magnetic field is in the y-z plane, and a mode with a wavelength perpendicular to the new mean magnetic field (\textit{i.e.} a wavelength in the x direction) would still be well resolved by this 1D simulation set-up (see Figure~\ref{fig:LamMirror}(a, right)).}
This reorientation of the magnetic field allows the development of magnetic mirror modes, a pressure-anisotropy-driven instability that occurs when 
\begin{equation}
\Lambda_\mathrm{mirror} = \sum_s \beta_{\perp,s}\left(\frac{P_{\perp, s}}{P_{\parallel, s}} - 1\right) > 1  \qquad, 
\end{equation}
where $s$ denotes the species and $\beta_{\perp,s} = 8 \pi n_s \chg{k_B} T_{\perp,s}/B^2$. 

Note that both the Bell and high-$\jcr$ cases develop significant pressure anisotropy prior to saturation because of ion cyclotron damping, but only the simulations in the high-$\jcr$ limit drive mirror modes on that timescale. 
This is illustrated in Figures~\ref{fig:LamMirror}(b) and (c) for the 1D runs in the Bell regime ($\jcr = 0.2$, black) and in the high-$\jcr$ regime ($\jcr = 4.0$, red). 
In both Figures~\ref{fig:LamMirror}(b) and(c) the vertical dashed line corresponds to the time at which the magnetic field saturates. 
For both cases, significant pressure anisotropy is present even before saturation, but because of the different growth rates, and the corresponding increased magnetic fields and $\beta_{\perp,s}$, only the high-$\jcr$ case is mirror-unstable ($\Lambda_\mathrm{mirror} > 1$) prior to saturation. 
Because this value can vary as a function of space, 
the maximum value of $\Lambda_\mathrm{mirror}$ over the simulation domain, $\Lambda_\mathrm{mirror}^\mathrm{max}$, is used to illustrate this point \citep{Ley2024}. 

The trend of only simulations in the high-$\jcr$ regime becoming mirror-unstable holds across the full range of $\jcr$ investigated in this work. This can be seen in Figure~\ref{fig:LamMirror}(d), where simulations with $\jcr \leq 2$ (to the left of the vertical, magenta, dashed line) are stable to the mirror instability at saturation, whereas every simulation in the high-current regime becomes mirror-unstable. 

The mirror modes that are growing in the high-$\jcr$ simulations may cause the instability to saturate through two potential channels: nonlinear interaction with the ion modes or particle trapping. 
The former is likely present to some extent as the expected range of wavenumbers over which the mirror modes should grow \citep{Gary1994} overlaps with the range of wavenumbers of the ion modes. 
The latter has been invoked as a possible explanation for the saturation of the two-stream instability \citep{davidson+72, McBride1972}. 
In these earlier papers, the particles were thought to be trapped within fluctuations driven by the two-stream instability itself, rather than mirror modes. 
While disentangling the extent to which each process contributes to the saturation is nontrivial given the highly nonlinear nature of both saturation mechanisms, the inability of the high-$\jcr$ instability to tap the full CR energy flux exactly when the mirror threshold is met strongly attests to mirror modes to be pivotal for saturation.
This finding is compatible with the simulations performed by \cite{Marret2022}, who noted that introducing a finite collisionality in kinetic simulations of the Bell instability may reduce pressure anistropies and eventually lead to a higher magnetic field at saturation.


\section{2D Simulations \label{sec:2D}}

\begin{figure}[ht!]
\includegraphics[width=\linewidth]{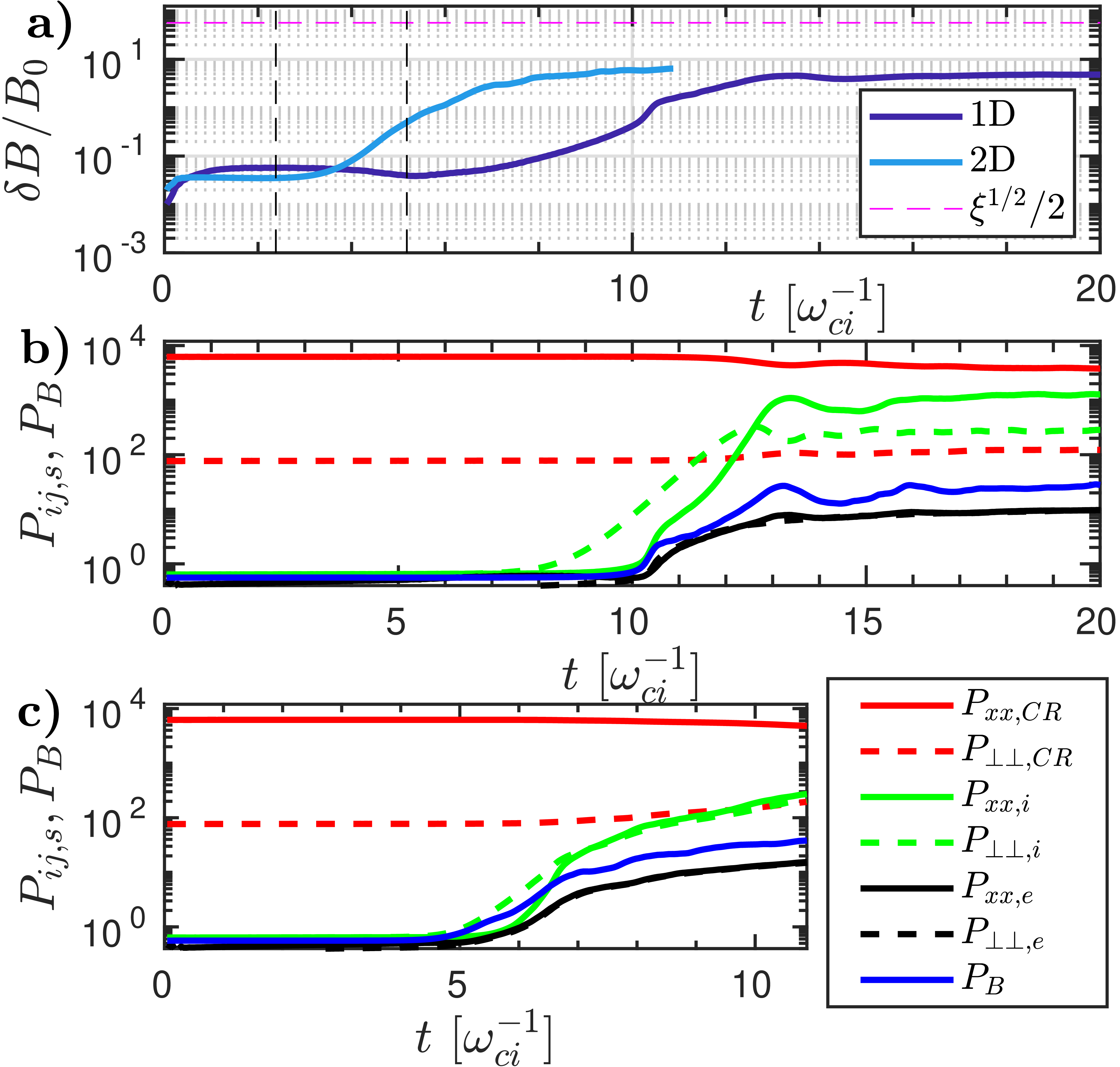}
\caption{\textbf{a)} $\delta B/B_0$ as a function of time for both the 1D (dark blue) and 2D (light blue) runs with $\tilde{J}_\mathbf{CR} = 10.0$ \chg{($\xi = 1.30 \times 10^4$)}. 
Components of the pressure tensor for \textbf{b)} the 1D and \textbf{c)} 2D simulations for $\jcr = 10.0$; 
for this 1D/2D study of modes with identical initial parameters, both the magnetic saturation and energy partition are remarkably similar.}
\label{fig:2DBaseline_Pt1}
\end{figure}

One natural question is whether the 1D setup is preventing further exchange of energy between CRs and other species, thereby artificially reducing saturation.
To check this, we performed some 2D simulations in the high-$\jcr$ regime. 
\cite{zacharegkas+24} have shown that the saturation of the Bell instability for parallel CR currents along $\mathbf{B}$ is effectively the same in 1D, 2D and 3D. 
To examine whether this holds in the high-$\jcr$ limit, we start by comparing two runs with the same parameters, $\jcr = 10.0$ and $\xi = 12600$, the overview of which is shown in Figure~\ref{fig:2DBaseline_Pt1}.

Overall the behavior in 1D and 2D simulations is similar, in particular for the major findings in this paper - the magnetic field saturation and the overall energy partition. 
The similarity in the magnetic field saturation level, controlled by the ion mode in 1D, can be seen in Figure~\ref{fig:2DBaseline_Pt1}(a), while the overall energy partition is illustrated in Figure~\ref{fig:2DBaseline_Pt1}(b) and (c). 

\begin{figure}[ht!]
\includegraphics[width=\linewidth]{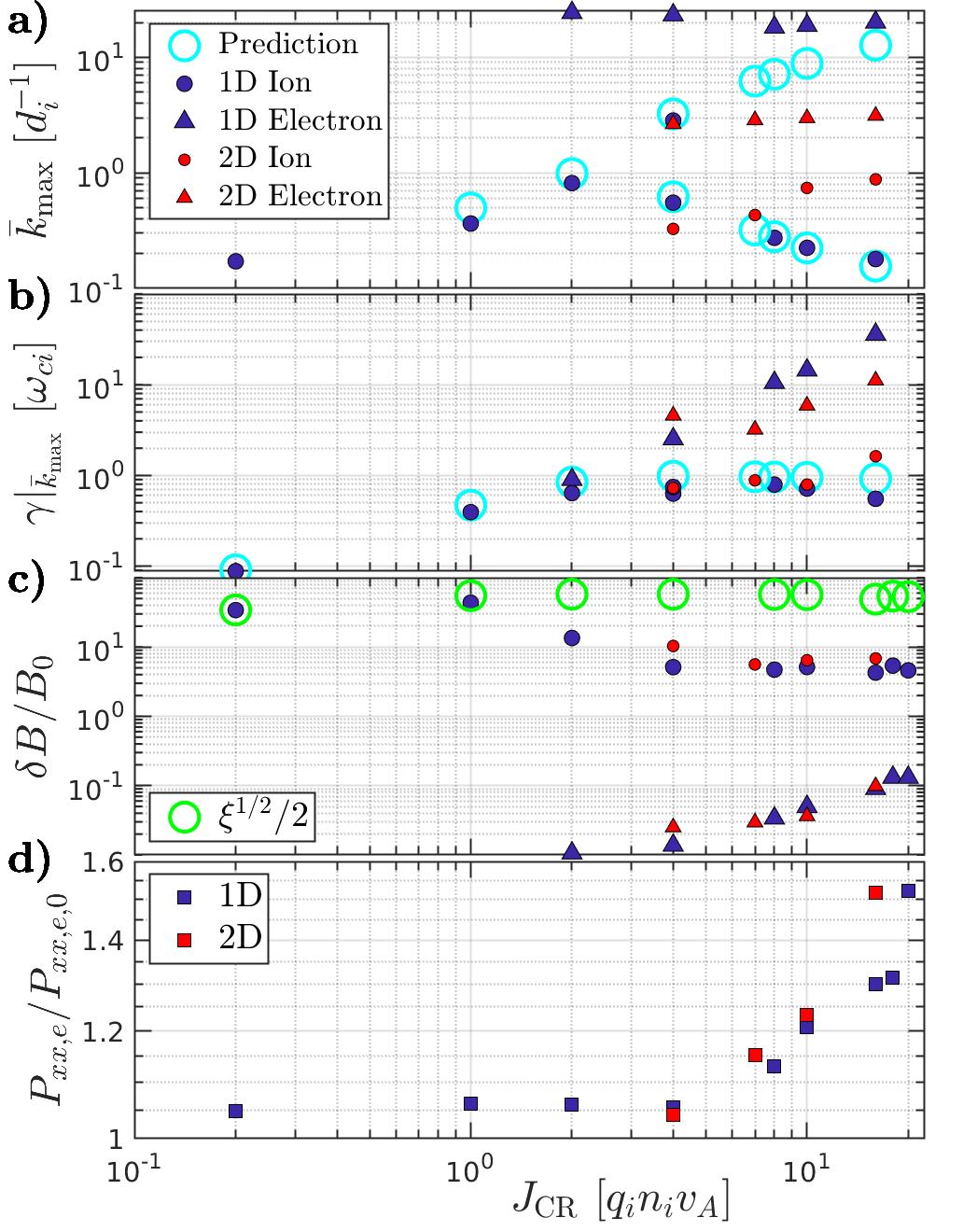}
\caption{\textbf{a)} ${k}_{max} d_i$ of the ion-scale (filled circles) and electron-scale (triangles) modes for both 1D (dark blue) and 2D (red) runs, compared with the prediction from Equation~\ref{eq:Weidl19} (open cyan circles). 
\textbf{b)} Measured growth rate for the same modes, compared with the corresponding predictions, as above. \textbf{c)} Transverse magnetic field fluctuations at saturation for electron modes ($t < \omega_{ci}^{-1}$) and ion modes ($t > \omega_{ci}^{-1}$) for 1D and 2D simulations. Open green circles correspond to the prediction $\delta B/B_0 = \xi^{1/2}/2$ \citep{zacharegkas+24}. 
\textbf{d)} Increase in the electron parallel pressure ($t < \omega_{ci}^{-1}$) for 1D and 2D simulations as a function of CR current, $J_\mathrm{CR}$. While the wavenumbers of the fastest growing mode are different between 1D and 2D, the consistency between the growth rate, magnetic field, and electron pressure suggest that most of the 2D physics is captured also in 1D.} \label{fig:Fig6_2DJcrScan}
\end{figure}

This consistency holds not just for the $\jcr = 10.0$ runs, but across a range of $\jcr$. 
In particular, the \chg{similarity} in the magnetic field saturation level between 1D and 2D can be seen in Figure~\ref{fig:Fig6_2DJcrScan}(c) for both the ion and electron mode saturation for the runs in Table \ref{tab:my_label1}. 
The ion thermal pressure exceeds both the magnetic field and electron pressure at saturation. 
Moreover, Figure~\ref{fig:Fig6_2DJcrScan}(d) shows that the change in the electron pressure due to the electron instability is nearly identical between the 1D and 2D runs. 
Throughout all the 2D runs there is a pressure anisotropy present throughout the simulation (Figure~\ref{fig:Fig6_BCycDamp}(c)) though at lower level than in the 1D runs, likely due to the increased difficulty in confining particles in 2D as compared to 1D.
The mirror criterion is met in both cases (Figure~\ref{fig:LamMirror}(d)), hence we conclude that the same saturation physics is at play.

For the modes themselves there are also marked similarities.
The growth rate of the fastest growing mode for both the ion and electron-scale modes match between the 1D and 2D runs within the intrinsic error in the estimation of the growth rate, as can be seen in Figure~\ref{fig:Fig6_2DJcrScan}(b). 
In particular for the electron-scale mode, because of the short timescales of the mode's evolution the growth rate estimation has higher intrinsic error. 
Given that the expected growth rate was derived for a 1D set-up \citep{weidl+19a}, this agreement is remarkable. 

The main difference between the 1D and 2D high-$\jcr$ simulations lies in the wavenumbers where the most unstable modes measured. 
This can be seen for both the ion and electron wavenumbers in Figure~\ref{fig:Fig6_2DJcrScan}(a). 
While the growth rate and saturation amplitude of the electron-scale mode are similar in 1D and 2D (Figure~\ref{fig:Fig6_2DJcrScan}b), Figure~\ref{fig:Fig6_2DJcrScan}(a) shows that the electron mode shifts by almost an order of magnitude in $k$, from \chg{an average $k_\mathrm{max,1D} = 18.8~d_i^{-1}$ to an average $k_\mathrm{max,2D} = 3.0~ d_i^{-1}$}, with the ion mode no longer exhibiting the decrease in wavenumber as a function of $\jcr$ in the high-$\jcr$ limit. 

\begin{figure}[ht!]
\includegraphics[width=\linewidth]{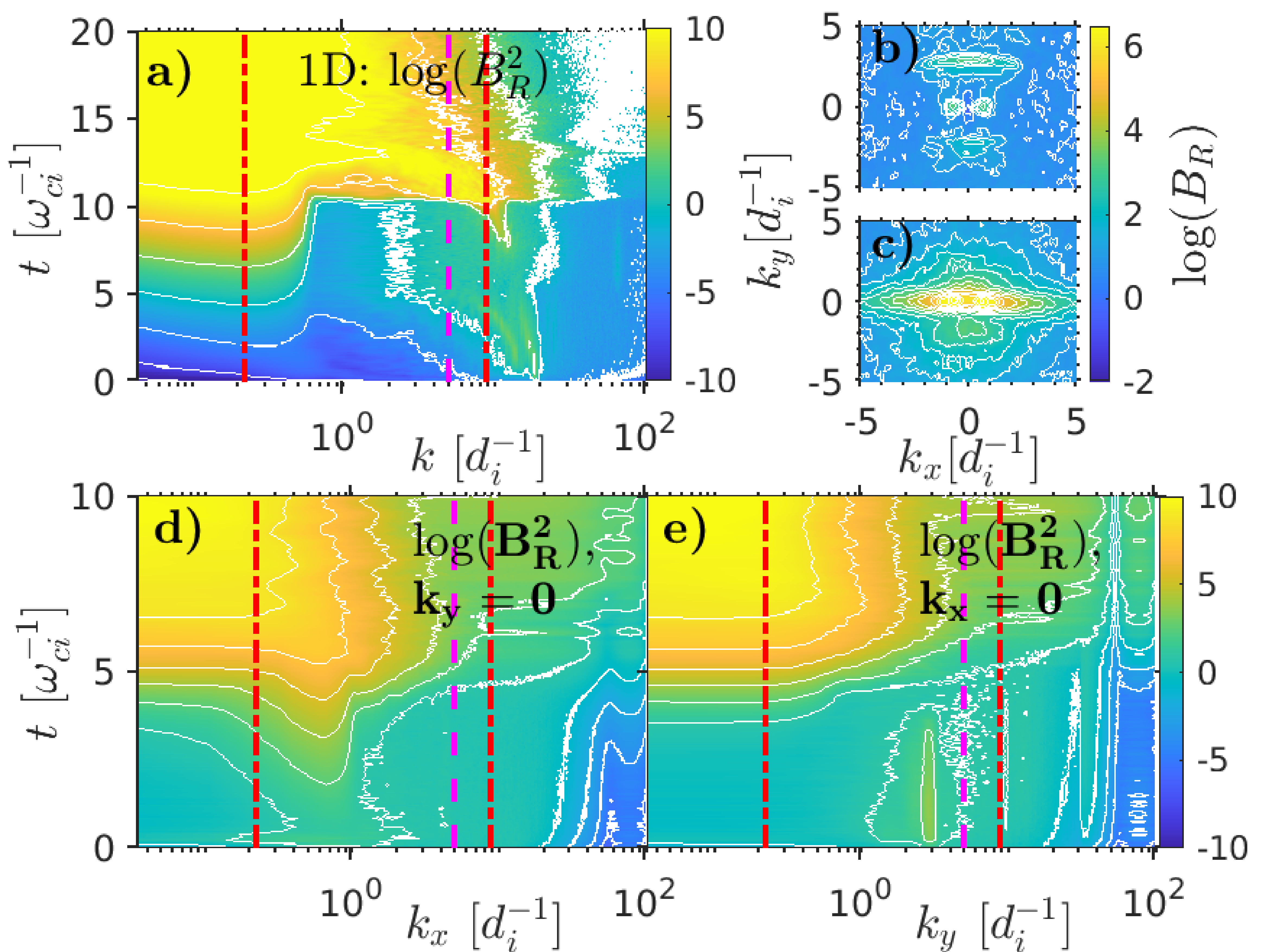}
\caption{\textbf{a)} Right-handed magnetic field amplitude as in Figure~\ref{fig:BaselineTSIFig2}(b) for the 1D simulation with $\jcr = 10$ \chg{($\xi = 1.30 \times 10^4$)} with the predictions from the high-$\jcr$ extension of Bell's instability, Equation~\ref{eq:kMaxBell} (pink, dashed) and Equation~\ref{eq:Weidl19}) (red, dot-dashed) overplotted. 
Slices of the right-handed B field, $|B_R|(k_x, k_y, t_i)$, for the 2D run at $t_i = 2.4 ~\omega_{ci}^{-1}$  (\textbf{b}), $t_i = 5.2~\omega_{ci}^{-1}$ (\textbf{c}),  $k_y = 0$ (\textbf{d}) and $k_x = 0$ (\textbf{e}). 
This case study at $\jcr = 10.0$ illustrates the main differences in the wavenumbers of the ion and electron modes between 1D and 2D - the magnitudes and the propagation direction of the electron mode.}
\label{fig:2DBaseline_Pt2}
\end{figure}

The properties of the 2D modes can be seen in more detail for the $\jcr = 10.0$ runs in Figure~\ref{fig:2DBaseline_Pt2}. 
Panels (b) and (c) show both the Fourier transform of the right-handed magnetic field, plotted as a function of $k_x$ and $k_y$ at a given time, while panels (d) and (e) show the magnetic field as a function of time and $k_x$ ($k_y$) at $k_y = 0$ ($k_x = 0$).
The time-points were chosen such that in (b) both the ion and electron modes were present and at the later time in (c) only the ion mode was present. 
The electron modes correspond to the arcs located roughly at $k_y d_i \sim \pm 2.5$, $k_x d_i \in [-2.5, 2.5]$ in Figure~\ref{fig:2DBaseline_Pt2}(b) and in Figure~\ref{fig:2DBaseline_Pt2}(e) at roughly $k_y d_i \sim 10^{0.2}$. 
The ion-scale modes correspond to the points at roughly $k_y d_i \sim 0$, $k_x d_i \sim \pm 0.8$ in Figure~\ref{fig:2DBaseline_Pt2}(b) and (c) and roughly $k_x d_i \sim 0.8$ in Figure~\ref{fig:2DBaseline_Pt2}(d).

\chg{We note that the smaller number of particles-per-cell used in our 2D simulations will affect the resolution of the small-wavelength modes. This effect can be seen in the dark blue artifact present in Figure~\ref{fig:2DBaseline_Pt2}(d) and (e) at roughly $k_{x,y}~d_i \sim 10^{1.4}$. This point is well-separated from the position of the 1D electron modes (which occur at roughly $k d_i \sim 10^{1.1})$, meaning the shift in electron mode is unlikely to be due to resolution effects. However wavenumbers past this point should be disregarded due to the impact of numerical effects.} 

With the additional dimensional freedom available in 2D, the electron-scale mode is now propagating perpendicular to the ion-scale mode, in a direction that was prevented by the 1D geometry. 
Additionally, while two ion modes were predicted in the 1D derivation from \citep{weidl+19a} and observed in the 1D simulations, in the 2D simulations there is no longer a second ion mode observed for $k_x d_i > 1$. While this mode was only observed without interference for the $\jcr = 4.0$ case in 1D, in 2D there is neither a second ion mode for $\jcr = 4.0$ or any sort of signature of interference where the 1D prediction of the second ion would be for $\jcr > 4$. 
Finally, we note that despite the same growth rate a different number of e-folds is required to achieve saturation in 1D and 2D (Figure~\ref{fig:2DBaseline_Pt1}a) due to the larger initial noise in the 2D run, which has fewer particles per cell to keep the run manageable.

While some of these differences are non-trivial, none of them substantially change either the main results (\textit{i.e.}, the lower-than-expected values of $B$ at saturation) or the physics that lead to the main result (\textit{i.e.}, the presence of ion-cyclotron damping that ultimately leads to the onset of the mirror instability and saturation).
We defer to a future work the study of the saturation in 3D, where there could be even more scattering and hence less anisotropy, though the difference between 1D and 2D suggests that the physics of saturation may not be profoundly different.  

\section{Conclusion\label{sec:conclusion}}
This work studied the behavior of the CR-driven non-resonant streaming instability in the limit of high CR current ($\jcr \ge 2$, i.e., for currents beyond the Bell regime) using PIC simulations. The main results of this investigation are as follows. 

\begin{itemize}
\item When the CR current is increased beyond $\jcr = 2$, the character of the ion mode driven by the relative drift between the CRs and the thermal ions fundamentally changes, splitting into two modes with the same growth rate capped at $\gamma \sim \omega_{ci}$ and with a substantial real frequency, consistent with the analytical predictions in \cite{weidl+19a}. 

\item Full PIC simulations reveal additional electron modes, which are fast-growing but saturate at small amplitude within one ion cyclotron time. 
As such, it is the ion modes that dominate the saturation of the plasma, setting the time scale on which the vast majority of the energy exchange and the magnetic field amplification occur. This suggests that even hybrid simulations, with fluid electrons, may be suited to capture the saturation of the high-$\jcr$ regime \citep[see, e.g.,][]{zacharegkas+19p}.

\item The electron mode typically contributes little to either the electron heating or the overall heating of the plasma. There is a special case in which the electron mode can efficiently heat the thermal electrons, namely when both the electrons and the CR beam are cold, in the sense that $v_d/v_{th,e} \gg 1$ and $p_{iso}/p_d \ll 1$. This case is discussed in detail, e.g., in \cite{matsukiyo+03, scholer+03, bohdan+20b, gupta+24a} in the context of the shock foot of supercritical non-relativistic shocks.

\item While the increase in the CR current corresponds to a stronger driving of the system, the high-current regime actually saturates well below the level extrapolated from the low-current regime. 
We see that in the high-$\jcr$ regime, the difference in the driven ion modes allows ion cyclotron damping to be more efficient. 
The resultant pressure anisotropy induced by such an ion-cyclotron heating drives mirror modes, which nonlinearly interact with and disrupt the ion modes, causing saturation of the instability below energy equipartition. 
Note that in the Bell regime, despite \chg{the fact that} cyclotron ion damping may be at work, the mirror threshold is never exceeded, which is optimal to extract a maximum field amplification.

\item These results hold in 2D as well, including growth rates, energization, and most critically, the saturation of the magnetic field below equipartition due to mirror modes. The main difference between 1D and 2D is in the wavenumber of the driven electron and ion modes. 
\end{itemize}

To summarize, we found that the high-$\jcr$ regime of the CR-driven instability is naturally prone to the  mirror instability driven by ion-cyclotron damping, which eventually has diminishing returns on the amount of magnetic field amplification provided by the CRs. 
These findings may be relevant for space and astrophysical situations in which strong currents in energetic particles can be produced. 
While these results hold in 2D, 3D simulations are needed to fully assess the actual level of anisotropy ensuing from ion cyclotron damping. 
 
\section*{ACKNOWLEDGMENTS}
The authors would like to thank K. Klein for assistance running \texttt{PLUME}. 
Simulations were performed on computational resources provided by the University of Chicago Research Computing Center and on TACC’s Stampede3 through ACCESS Maximize allocation PHY240042.
This work was partially supported by NASA through grants 80NSSC18K1218, NSSC23K0088, and 80NSSC24K0173 and NSF through grants PHY-2010240 and AST-2009326.
E.~L. was also partially supported by NSF Award No. 1949802 and by NRL base funding and the 2024-2035 Karle Fellowship. This work was performed in part at the Aspen Center for Physics, which is supported by National Science Foundation grant PHY-2210452.


\bibliography{Total,  ElecEn2023}
\bibliographystyle{aasjournal}

\end{document}